\documentclass[sigconf]{acmart}
\usepackage{balance}
\usepackage{bm}
\usepackage{adjustbox}
\usepackage{graphicx,xcolor} 
\usepackage[flushleft]{threeparttable}
\usepackage{balance}
\usepackage{algorithm}
\usepackage[noend]{algpseudocode}
\usepackage{multirow}
\usepackage{tabularx,booktabs}
\usepackage{listings}
\usepackage{xcolor}
\usepackage{amsmath}

\algblock{ParFor}{EndParFor}
\algnewcommand\algorithmicparfor{\textbf{for}}
\algnewcommand\algorithmicpardo{\textbf{parallel do}}
\algnewcommand\algorithmicendparfor{\textbf{end\ parallel for}}
\algrenewtext{ParFor}[1]{\algorithmicparfor\ #1\ \algorithmicpardo}
\algrenewtext{EndParFor}{\algorithmicendparfor}
\AtBeginDocument{%
  }

\setcopyright{acmlicensed}
\copyrightyear{2018}
\acmYear{2018}
\acmDOI{xxx.xxx}

\acmConference[Conference acronym 'XX]{Make sure to enter the correct
  conference title from your rights confirmation emai}{June 03--05,
  2018}{Woodstock, NY}
\acmISBN{978-1-4503-XXXX-X/18/06}

\begin{document}


\title{APEX: An Extensible and Dynamism-Aware Simulator for Automated Parallel Execution in LLM Serving}

\author{Yi-Chien Lin}
\email{yichienl@usc.edu}
\affiliation{%
  \institution{University of Southern California}
  \city{Los Angeles}
  \state{California}
  \country{USA}
}

\author{Woosuk Kwon}
\email{woosuk.kwon@berkeley.edu}
\affiliation{%
  \institution{University of California, Berkeley}
  \city{Berkeley}
  \state{California}
  \country{USA}
}

\author{Ronald Pineda}
\email{ronaldp10@ucla.edu}
\affiliation{%
  \institution{University of California, Los Angeles}
  \city{Los Angeles}
  \state{California}
  \country{USA}
}

\author{Fanny Nina Paravecino}
\email{fanny.nina@microsoft.com}
\affiliation{%
  \institution{Microsoft}
  \city{Mountain View}
  \state{California}
  \country{USA}
}

\renewcommand{\shortauthors}{Lin et al.}

\begin{abstract}
Efficiently serving Large Language Models (LLMs) requires selecting an optimal parallel execution plan, balancing computation, memory, and communication overhead. 
However, determining the best strategy is challenging due to varying parallelism techniques (data, pipeline, tensor) and workload characteristics (e.g., compute-intensive tasks with long prompts vs. memory-intensive tasks with long generation).
We propose APEX, an LLM serving system simulator that efficiently identifies optimal parallel execution plans by considering key factors of LLM serving systems, such as memory usage, batching behavior, etc.
APEX performs dynamism-aware simulation to model iteration-level batching, and leverages LLMs' repetitive structure to reduce design space, scaling efficiently to trillion-scale models. 
APEX abstracts the key components of LLM serving systems—including the model, batching module, quantization formats, and device clusters—enabling the simulator to be general and extensible.
Simulating on a CPU, APEX evaluates execution plans for various device clusters, covering diverse LLMs and workloads. 
APEX finds plans up to 3.37$\times$ faster than heuristics, and also plans that reduce energy consumption by up to 45\% compared to latency-optimal plans.
APEX performs comprehensive evaluations, reporting key system metrics like time per output token and time to first token, which can help service providers meet SLOs.
APEX identifies an optimal plan within 15 minutes on a CPU, making it 71$\times$ faster and 1234$\times$ more cost-effective than cloud-based GPU deployment.
APEX can be accessed at: \url{https://github.com/microsoft/apex_plus}
\end{abstract}




\settopmatter{printfolios=true}
\maketitle

\section{Introduction}\label{sec:intro}

Large Language Models (LLMs) have been successfully applied to various applications, such as code generation \cite{code_gen1,code_gen2}, question-answering \cite{qa1,qa2}, and many more \cite{llm4eda,llm4rec,llm4telcomm}. 
Serving an LLM is both memory- and compute-intensive, necessitating multi-device clusters to achieve high performance.
Several techniques have been proposed to parallelize LLM, such as data parallelism (DP) \cite{dp}, pipeline parallelism (PP) \cite{gpipe}, tensor parallelism (TP) \cite{megatronlm}, etc.
Each parallelism has its pros and cons: TP is memory efficient, as it does not produce model replicas like DP, but it incurs an expensive communication overhead \cite{megatronlm}.
PP reduces communication overhead, but can suffer from workload imbalance \cite{gpipe}.
Hybrid parallelism can effectively balance the trade-offs to improve performance, but identifying the optimal parallel execution plan is non-trivial due to the many factors involved, including computation workload, network traffic, memory efficiency, etc.
Furthermore, the optimal parallel execution plan also depends on the characteristics of the input requests. 
Compute-intensive tasks (e.g., tasks with long prompts like summarization) and memory-intensive tasks (e.g., tasks with long generations like coding) favor different parallelism strategies. 
A common practice is to adopt heuristic execution plans instead, such as applying TP within the same node and PP across different nodes \cite{flexflow,heuristic}.
Yet, recent studies show that heuristics plans can be 2$\times$ slower than optimal configurations \cite{calculon}, underscoring the need for service providers to search for and adopt an optimal parallel execution plan rather than relying on heuristics.

To find an optimal execution plan, a straightforward way is to deploy and evaluate parallel execution plans exhaustively; however, this is prohibitively expensive as it would take thousands of GPU hours \cite{Sarathi-Serve,vidur}.
An alternative is to leverage modeling-based approaches, such as performance models \cite{calculon} or simulators \cite{habitat,proteus}, to estimate the performance of  parallel execution plans;
nevertheless, existing solutions developed for Deep Neural Network \cite{habitat,proteus} or LLM training \cite{calculon,vTrain} cannot be applied to LLM serving, as serving systems introduce unique challenges:
\textbf{(1) Dynamism of iteration-level batching:}
Unlike conventional ML systems that use static batching, where all requests in a batch must be completed before new ones are added, LLM serving systems \cite{vllm,Sarathi-Serve,distserve} employ \textit{iteration-level batching} \cite{orca}. 
In this approach, incoming requests are continuously added to the processing batch as memory becomes available, resulting in dynamically changing batch sizes which are hard to model.
Moreover, some requests within the batch may be in the prefill stage, while others are in the decode stage. 
This interleaving of stages further complicates the modeling, as these stages have significantly different computational characteristics (details in Section \ref{sec:bg_inference}).
\textbf{(2) Exponentially-growing design space:}
The design space grows exponentially with respect to the model size and the number of devices in the cluster.
Since LLM serving often involves large models and clusters, it is non-trivial to develop a solution that can search for an optimal parallel execution plan within a reasonable timeframe.
\textbf{(3) Adapting for continuously-evolving systems:}
LLM serving system is an emerging area, with continuous evolution in model architecture, hardware platforms, and system optimizations such as quantization \cite{awq,kvquant} and parallelism techniques \cite{expert_par,context_par}. 
Consequently, a performance model or a simulator can easily become obsolete and inapplicable to the latest LLM serving systems. 
Thus, keeping pace with the rapid evolution of such systems is necessary but also challenging.

To this end, we propose APEX, an extensible and dynamism-aware simulator for
\underline{A}utomated \underline{P}arallel \underline{EX}ecution in LLM serving.
Given an LLM, a device cluster, and input requests with varying context/generation lengths, APEX generates an optimal parallel execution plan for LLM serving.
APEX first generates various parallel execution plans, each representing a unique way to parallelize the model by combining various parallelisms (i.e., DP, PP, TP).
APEX then evaluates each plan by estimating the execution time of serving the input requests through simulation. 
APEX performs dynamism-aware simulation that is capable of modeling the characteristics of iteration-level batching.
After simulation, APEX provides a comprehensive evaluation for each execution plan with key LLM serving metrics, such as time per output token (TPOT), time to first token (TTFT), and energy consumption, among others \cite{LLM_survey}.
This detailed evaluation enables service providers to select execution plans optimized for various objectives (e.g., minimizing latency or energy consumption) while meeting service-level objectives (SLOs).

Despite performing complex simulations, APEX remains highly scalable by leveraging the repetitive structure of transformer models, significantly reducing design space and enabling simulations at trillion-scale models across multi-node clusters. 
To model state-of-the-art LLM serving systems, APEX supports a broad range of LLMs, parallelisms, quantization formats, and device clusters.
Recognizing the rapid evolution of LLM serving, APEX is designed to be modular and extensible, allowing easy integration of new models and hardware with minimal effort.
Additionally, APEX leverages operation-level profilings to estimate the system execution time, which allows APEX to adapt to different clusters by collecting performance data from the underlying platform. 
Note that APEX not only supports GPU clusters but also ASIC clusters such as TPU \cite{TPUv4} and Intel Gaudi \cite{gaudi}.
We evaluate APEX using a myriad of LLMs, device clusters, and distinct workloads. APEX consistently identifies optimal execution plans that outperform heuristic plans while maintaining low simulation overhead. APEX is open-sourced:  {\url{https://github.com/microsoft/apex_plus}}.
Our key contributions are:
\begin{itemize}
    \item \textbf{Automated parallel execution}: We propose APEX, a simulator that automatically finds an optimal parallel execution plan for a given LLM, a device cluster, and input requests.
    \item \textbf{Dynamism-aware simulation}: APEX models iteration-level batching, capturing real-world LLM serving behavior.
    \item \textbf{Accurate \& effective}:  APEX delivers accurate speedup predictions, with an average relative error of just 10.7\%. APEX finds optimal plans up to 3.37$\times$ faster than heuristic plans, and energy-optimal plans that reduce energy consumption by up to 45\% compared to latency-optimal plans.
    \item \textbf{Broad compatibility and extensibility}: APEX supports various LLMs, parallelisms, quantizations, and device clusters, while allowing easy extension to new configurations.
    \item \textbf{Cost-effective \& scalable}: APEX identifies an optimal plan within 15 minutes on a CPU, making it 71$\times$ faster and 1234$\times$ more cost-effective than cloud-based GPU deployment. It also maintains similar simulation overhead when scaling from billion-scale to trillion-scale models.
    \item \textbf{Comprehensive evaluation}: APEX reports key LLM-serving metrics, which can help service providers meet SLOs.
\end{itemize}

\section{Background}\label{sec:background}
\subsection{LLM Inference}\label{sec:bg_inference}
Large Language Model (LLM) inference takes a user prompt as input, and generates a response token by token.
The sequential token generation process is also known as auto-regressive generation~\cite{autoregressive}, which utilizes previously generated tokens as input to predict the next token.
The inference process of LLMs consists of two stages:

\noindent \textbf{1. Prefill Stage:}
In the prefill stage, LLM processes the input request (i.e., prompt) to set up intermediate states (keys and values) that are used to predict the first token. Unlike token generation, the computation of prefill does not rely on previously generated output tokens, allowing the tokens of the input request to be processed in parallel all at once. 
This high degree of parallelism makes the prefill stage compute-bound \cite{distserve}.

\noindent \textbf{2. Decode Stage:}
During the decode stage, LLM generates output tokens autoregressively until a stopping criterion is met. 
The generation of each token depends on all previous tokens' output states (keys and values). 
The generation phase is often memory-bound \cite{awq}, as its latency is primarily determined by the data transfer time of reading the output states from memory.

\subsection{Device Clusters and Interconnections}
\begin{figure}[t]
    \centering
    \includegraphics[width=6.2cm]{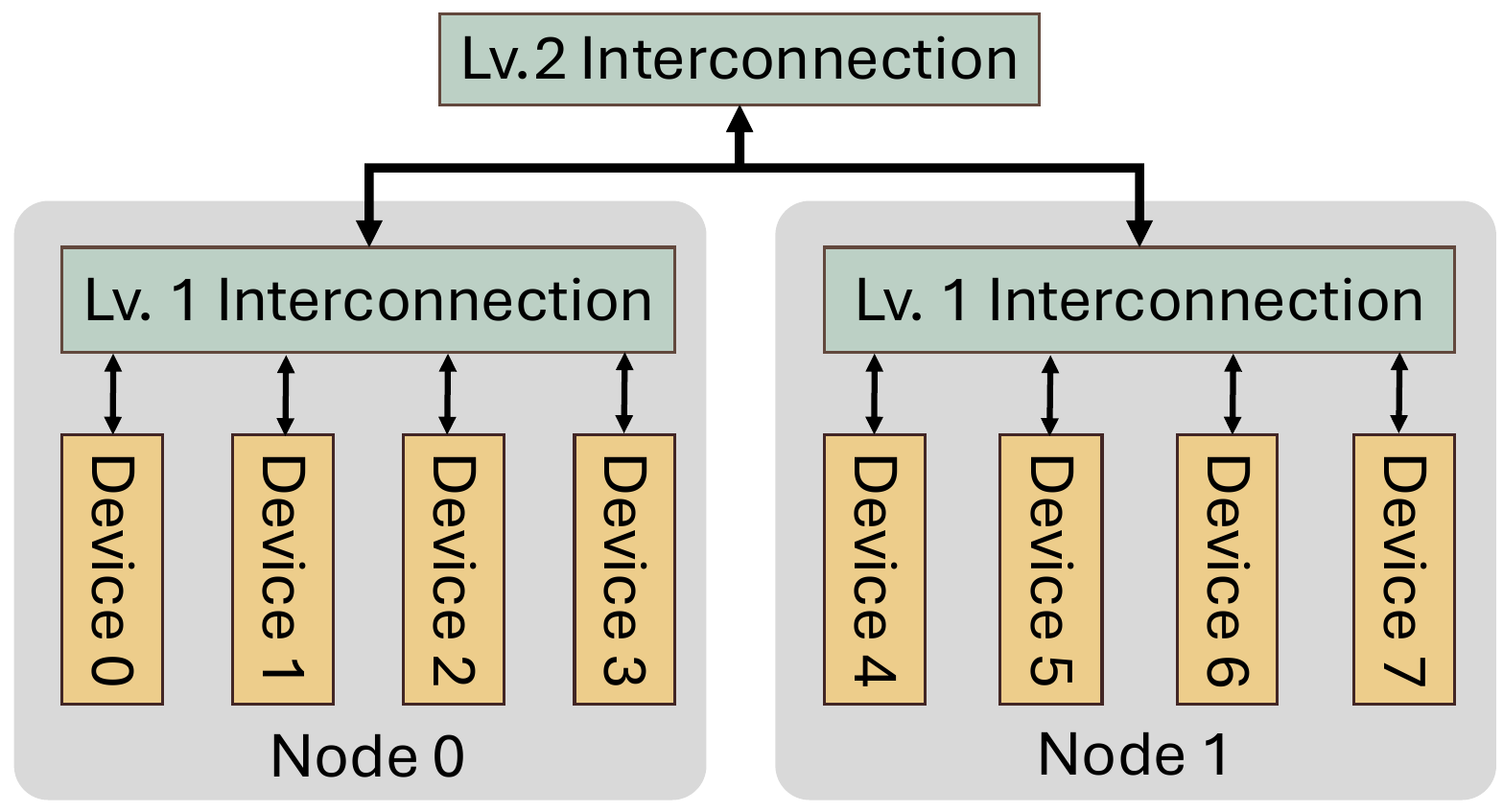}
    \caption{An example two-level device cluster. Memory bandwidth and latency are uniform within the same level.}
    \vspace{-0.7cm}
     \label{fig:cluster}
\end{figure}

To achieve high serving throughput and accommodate the large model size, LLMs are often deployed on a cluster of devices.
The devices are connected in a tree-based network topology, which is one of the most popular topologies for node deployment.
The memory bandwidth and latency is uniform within each level.
Figure \ref{fig:cluster} illustrates an example of a commonly used two-level device cluster.
Devices within the same node are typically connected by high-bandwidth interconnects like NVLink \cite{gpu_connect} at level 1, while devices across different nodes are connected via inter-node networks such as InfiniBand \cite{infiniband} at level 2. 
In addition to GPUs, ASIC clusters are also emerging, such as TPU and Gaudi clusters \cite{TPUv4,gaudi}. 
These clusters also utilize tree-based network topologies for interconnection and can be abstracted similarly as GPU clusters \cite{h100}.

\subsection{LLM Serving and Iteration-Level Batching}\label{sec:serving}
LLMs are often hosted by service providers in the cloud \cite{llm_serving}. 
Users send requests to these providers through APIs or chatbots and then receive responses generated by the LLMs.
For LLM serving, it is critical to achieve both low latency and high throughput: low latency is necessary to meet service-level objectives, while high throughput enables service providers to serve a large number of users simultaneously.
Due to the auto-regressive nature of LLMs, the lengths of the generated responses can vary significantly.
Consequently, the commonly used static batching leads to suboptimal performance in LLM serving, as all the batched requests need to wait for the longest response to be generated to proceed to batch new requests. 
To overcome this inefficiency, \textit{iteration-level batching} \cite{orca} is proposed.
Iteration-level batching continuously schedules newly arrived request(s) into the existing batch whenever GPU memory becomes available, rather than waiting for all the requests to be completed.
Iteration-level batching significantly improves the serving throughput and is widely adopted in state-of-the-art LLM serving systems \cite{vllm,Sarathi-Serve,distserve,tensorRT}.

\subsection{LLM Parallelisms}\label{sec:parallel}
LLM is often served with multiple devices (Figure \ref{fig:cluster}), and various approaches have been proposed to parallelize LLM on the cluster. We discuss several representative and widely-used parallelisms.

\noindent \textbf{Data Parallelism (DP)}: In DP \cite{dp}, the model is replicated across multiple devices, and input requests are divided into mini-batches, each processed independently by a replica. This approach avoids communication overhead during execution, but comes at the cost of increased memory usage due to model replication, thereby reducing available memory for batching requests.

\noindent \textbf{Pipeline Parallelism (PP)}: PP \cite{gpipe,pipedream} divides the model in a layer-wise fashion. Each model partition consists of a subset of layers, and each subset of layers forms a pipeline stage. The input requests are split into micro-batches to flow through the pipeline stages. PP requires point-to-point (p2p) communications between the pipeline stages. In addition, PP suffers from pipeline stalls when the execution time of each stage is unbalanced.

\noindent \textbf{Tensor Parallelism (TP)}: TP \cite{megatronlm} divides the model in an intra-layer fashion, which splits individual layers across multiple devices. TP does not suffer from load imbalance like PP or produces model replicas like DP. Yet, TP incurs high collective communication overhead, such as AllReduce, which is prohibitively expensive for interconnection networks with limited bandwidth like PCIe.

\noindent \textbf{Expert Parallelism (EP)}: EP \cite{expert_par} is a special type of parallelism for Mixture of Experts (MoE) models~\cite{moe}. EP deploys different experts across various devices. For MoE models, only a subset of experts are activated for each input token. Thus, EP can result in workload imbalance and resource under-utilization if none of the experts on the device are activated; while this can be avoided by adopting TP, where each device is assigned with a partition of all the experts, EP incurs a lower communication overhead than TP.

Each parallelism has its own pros and cons, tradeing-off between computation, memory efficiency, collective communication overhead, etc. Thus, it's non-trivial to determine the optimal parallel execution plan, especially given that the parallelisms can be adopted in a hybrid manner.
Heuristic plans are typically adopted for simplicity, such as applying TP to clusters with high-bandwidth interconnections like NVLink \cite{gpu_connect}, and applying PP to clusters without such networks \cite{flexflow,heuristic}. 
APEX aims to evaluate the complex trade-offs among parallelism techniques to find optimal parallel execution plans that outperform heuristic approaches.

\subsection{LLM Quantizations}
Quantization is an essential technique for achieving performant LLM serving, as it reduces both memory and computation overhead.  
Various methods have been proposed to quantize one or more of the following components to lower precision while preserving high accuracy: (1) model weights, (2) activations, and (3) KV cache.
For example, AWQ \cite{awq} quantizes the model weights, and KVQuant \cite{awq} focuses on quantizing the KV cache.
State-of-the-art LLM serving systems like vLLM \cite{vllm} and TensorRT-LLM \cite{tensorRT} also support W8A8 quantization, which quantizes both the model weights and activations to FP8 format.
Given the diversity of quantization methods, it is essential to flexibly support various techniques to find an optimal parallel execution plan for LLM serving systems.

\section{APEX Simulator Design}\label{sec:system}
\begin{figure}[t]
    \centering
    \includegraphics[width=8cm]{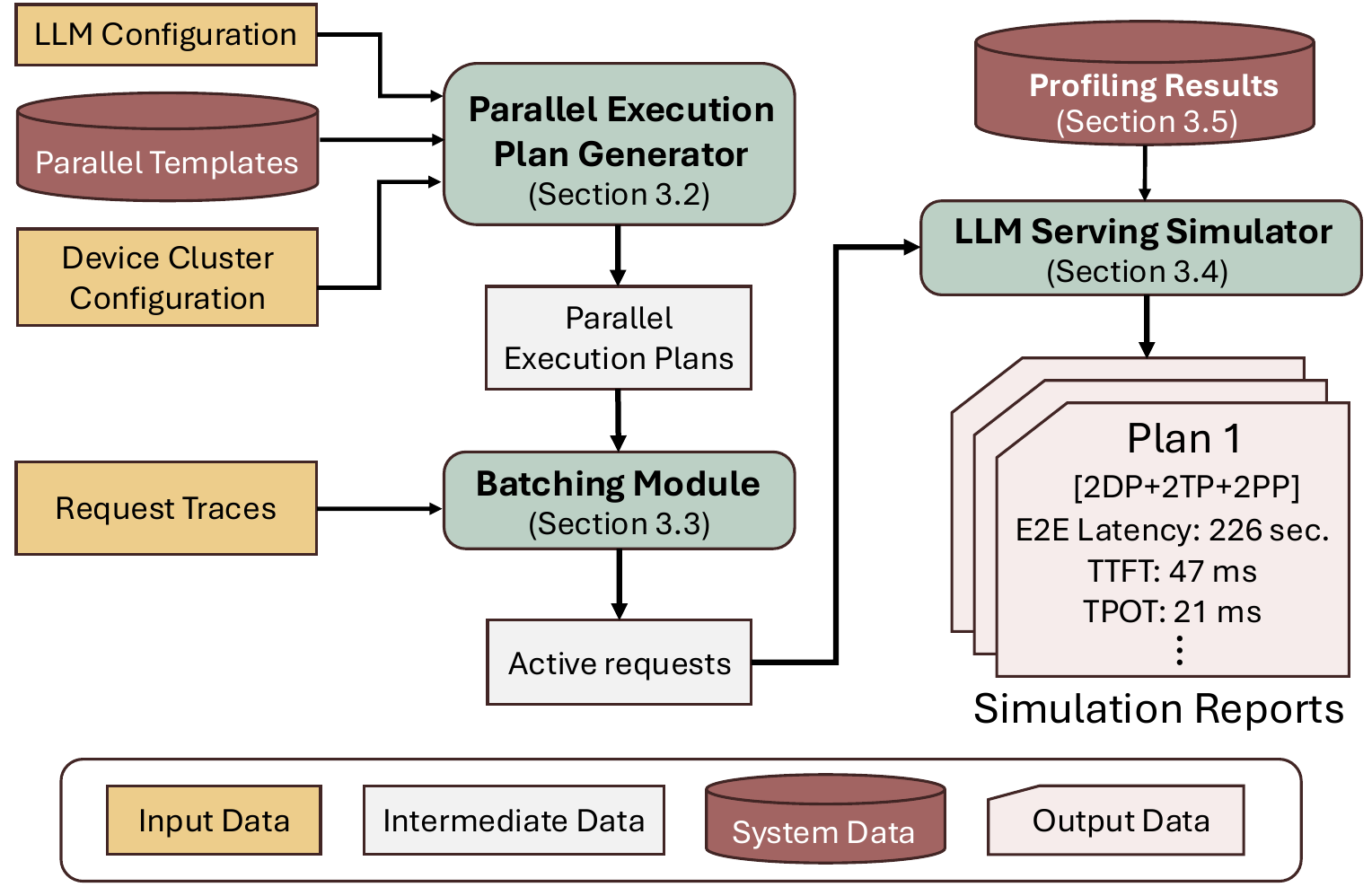}
     \vspace{-0.35cm}
    \caption{System overview of APEX}
\vspace{-0.3cm}
     \label{fig:overview}
\end{figure}

\subsection{System Overview}
We depict the workflow of APEX in Figure \ref{fig:overview}. 
User-provided input data is shown in yellow-colored boxes, and built-in system data is represented by red-colored cylinders.
Given an LLM and a device cluster, APEX generates multiple parallel execution plans using built-in Parallel Templates. 
Each generated executon plan represents a distinct strategy for mapping the LLM across the cluster. 
The Batching Module takes these generated plans and a trace of requests as inputs, dynamically generating the processing batch based on request arrival time and resource constraints. 
Specifically, the Batching Module determines, at each iteration, whether a request should be added to or removed from the processing batch and reports the active requests to the LLM Serving Simulator;
the Simulator then uses profiling results to estimate execution time and energy consumption for processing the active requests, while also tracking system metrics such as time to first token (TTFT), time per output token (TPOT), P95 latency, among others. 
Once all requests are processed, the simulator generates a simulation report for each parallel execution plan. 
APEX can optimize towards different objectives by selecting the most suitable execution plan based on a parametrizable target metric:
It can optimize for energy-efficiency by minimizing energy per inference, or can prioritize throughput by reducing end-to-end latency, enabling flexible adaptation to diverse system goals.
Suppose that we aim to find the latency-optimal plan for serving Llama-3.1 70B on a cluster with 8 Nvidia H100 GPUs. 
APEX first generates multiple execution plans, such as 8-way tensor parallelism, or a 4-way tensor parallelism combined with 2-way pipeline parallelism.
Next, it utilizes a trace of requests as input for simulation, such as the Azure LLM Traces \cite{splitwise}.
APEX simulates iteration-level batching to organize arriving requests into the processing batch. 
Finally, the LLM Serving Simulator estimates the execution time of each plan and select the one with the lowest end-to-end latency for serving Llama-3.1 70B on the cluster.
\begin{figure}[t]
    \centering
    \includegraphics[width=7.6cm]{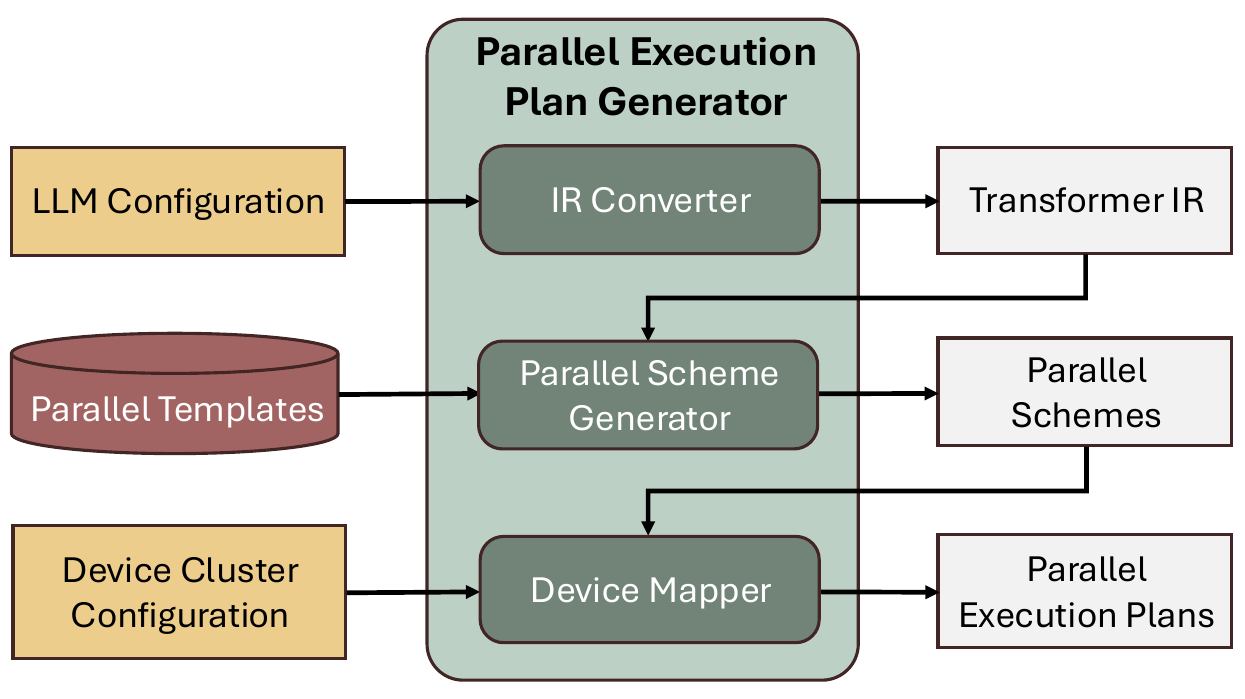}
    \caption{The Parallel Execution Plan Generator produces plans that map a given LLM onto the target cluster using distinct parallelization strategies}
    \vspace{-0.5cm}
     \label{fig:plan_gen}
\end{figure}

\subsection{Parallel Execution Plan Generator}
In this section, we describe the details of the Parallel Execution Plan Generator, which constructs various parallel execution plans to map the input LLM onto the target device cluster.
The workflow of the Plan Generator is illustrated in Figure \ref{fig:plan_gen}.
The process begins by converting the input LLM into Transformer IR, an intermediate representation that provides a canonical abstraction for transformer-based models.
Using this representation, the Plan Generator produces multiple \textit{parallel schemes} based on the Parallel Templates.
A \textit{parallel scheme} maps the model onto a \textit{logical device cluster} \cite{hiveD}, a virtual overlay that abstracts hardware details without network topology information.
Next, the Device Mapper assigns logical devices to {physical devices}, generating executable \textit{parallel execution plans}, which serve as inputs for the Batching Module.
This two-stage mapping approach (first to the logical and then to physical device cluster) reduces the design complexity.
We discuss each module in detail below.

\subsubsection{IR Converter and Transformer IR}\label{sec:ir}
\begin{figure}[t]
    \centering
    \includegraphics[width=7.5cm]{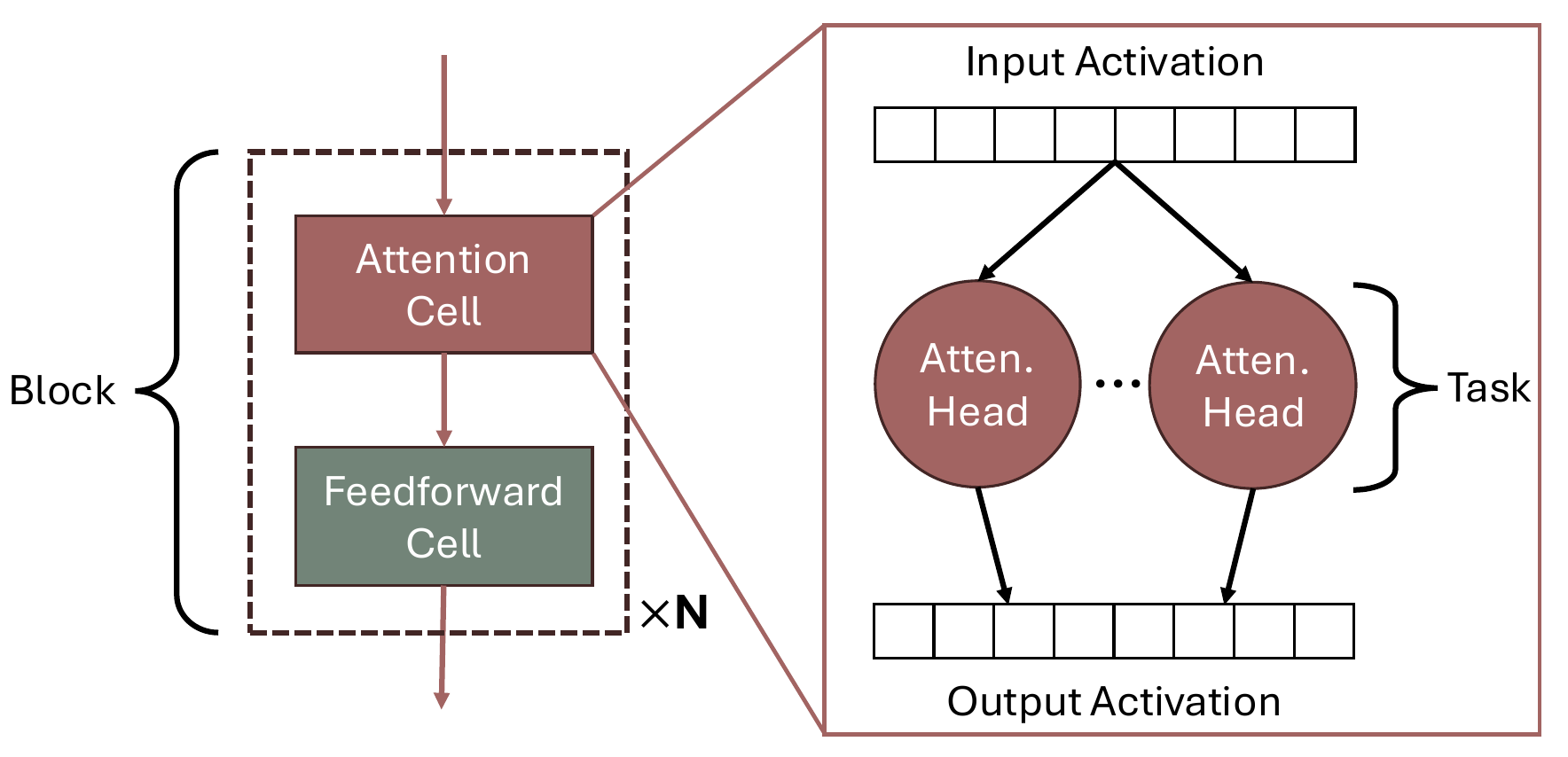}
    \vspace{-0.2cm}
    \caption{Transformer IR represents LLMs in a canonical way}
    \vspace{-0.3cm}
     \label{fig:IR}
\end{figure}

The Plan Generator converts an LLM into Transformer IR, a canonical representation of transformer-based models, as illustrated in Figure \ref{fig:IR}. We develop an IR converter that parses an LLM’s configuration file, extracting parameters such as the number of attention heads and layers, and translates the model into Transformer IR.
In Transformer IR, an LLM is represented as a chain of \textit{cells}, where each cell corresponds to key transformer operations, such as multi-head attention or the feedforward network. 
For example, GPT-3 \cite{gpt3} can be represented as a sequence of multi-head attention (MHA) cells and multi-layer perceptron (MLP) cells, while Llama-3.1 \cite{llama3} is modeled as a chain of group query attention (GQA) cells \cite{gqa} and SwiGLU cells \cite{swiglu}.
Each cell consists of multiple \textit{tasks}, such as individual attention heads in an MHA cell or experts in a mixture-of-experts (MoE) cell. 
Given that LLMs typically consist of repeated structures, we define the smallest set of nonrepetitive adjacent cells as a \textit{block}. 
Using Transformer IR, an LLM is uniformly represented as multiple identical \textit{block}, where each block consists of \textit{cells}, and each cell contains multiple \textit{tasks}.
Transformer IR abstracts an LLM by focusing solely on key operations like MHA and MLP, while omitting operations like tokenization and position embeddings, which are less relevant for model parallelization. 
By focusing on key operations and leveraging the repetitive nature of LLMs through \textit{blocks}, Transformer IR reduces both the search space and the simulation runtime for APEX. Furthermore, Transformer IR serves as a foundation for developing our Parallel Templates, which we describe in the following subsection.


\subsubsection{Parallel Templates and Parallel Schemes Generator}\label{sec:template}
\begin{figure}[t]
    \centering
    \includegraphics[width=8.5cm]{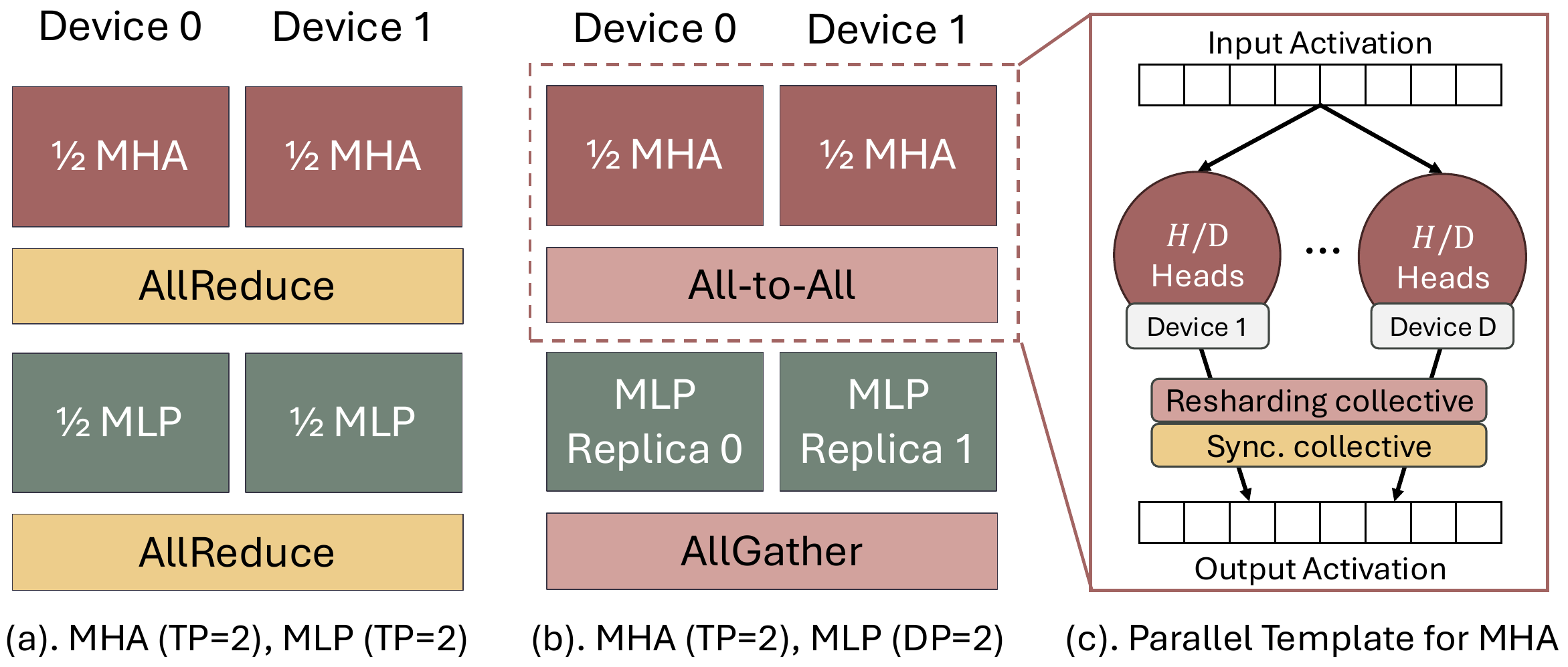}
    \caption{Examples of APEX's Parallel Templates for two devices. The parameterized templates can extend to \textit{D} devices.}
    \vspace{-0.3cm}
     \label{fig:template}
\end{figure}

We develop Parallel Templates to generate parallel execution plans for LLM serving. 
Given the wide variety of LLMs, exhaustively designing templates for each model is impractical. 
Instead, we develop our Parallel Templates based on the Transformer IR, which provides a unified abstraction of LLMs. 
This approach enables the developed templates to support a broad range of models rather than being tailored to specific ones.
Figure \ref{fig:template} illustrates examples of our templates. 
Each \textit{cell} type is associated with a pre-defined template that specifies how it is parallelized across multiple devices. 
For instance, if tensor parallelism (TP) is applied to both the multi-head attention (MHA) cell and the multi-layer perceptron (MLP) cell, as shown in Figure \ref{fig:template} (a), \textit{tasks} (e.g., attention heads) are evenly distributed across devices.
Beyond task distribution, Parallel Templates also manage collective communication between adjacent cells. 
In Figure \ref{fig:template} (a), an AllReduce operation synchronizes data between cells, similar to the approach in Megatron-LM \cite{megatronlm}. 
In Figure \ref{fig:template} (b), 2-way data parallelism (DP) is applied to the MLP cell, while no DP is applied to the MHA cell, leading to distinct model partitioning. 
Consequently, All-to-All and AllGather operations are performed for tensor resharding.
While Figures \ref{fig:template} (a) and (b) depict simplified examples using two devices, Parallel Templates are fully parameterized to support any number of devices, 
as in Figure \ref{fig:template} (c), where $H$ attention heads are evenly distributed across $D$ devices.


Given Parallel Templates and an LLM represented in Transformer IR, the Parallel Scheme Generator constructs parallel schemes that map the LLM onto a \textit{logical device cluster} \cite{hiveD} using different combinations of parallelism techniques. 
A logical device cluster is a virtual overlay that specifies the number of devices in a cluster without considering network topology. 
As a result, a parallel scheme defines the parallelism applied to Transformer IR cells and their collective communications (e.g., Figure \ref{fig:template} (a), (b)) without assigning the cells to specific physical devices.
Introducing logical device clusters simplifies design complexity by decoupling Parallel Templates from the hierarchical network topology of physical clusters; the Device Mapper later handles physical mappings. 
The Parallel Scheme Generator follows a hierarchical top-down approach, starting with coarse-grained model-level parallelism and progressively refining to fine-grained cell-level parallelism. 
As detailed in Algorithm \ref{alg:parallel_scheme}, the Generator first selects the degree of model-level data parallelism (i.e., number of model replicas) and partitions the cluster accordingly. 
APEX enforces even partitioning, restricting parallelism choices to divisors of the total number of devices. 
Next, the Generator determines the number of pipeline stages within each replica, again ensuring even partitioning. 
Finally, it assigns cell-level data parallelism and applies Parallel Templates to generate task mappings.
If a cell's data parallelism degree is lower than the number of assigned devices, intra-layer parallelism (e.g., tensor or expert parallelism) is applied. 
For instance, if four devices are assigned to a cell with two replicas, each replica receives two devices, and tensor parallelism is applied to parallelize the cell replica across the two devices.
Lastly, for each pair of adjacent cells, the Generator inserts collective communication operations for synchronization and tensor resharding.
Now that various parallel schemes have been generated for the target model, the logical devices can be mapped to physical devices, producing executable parallel execution plans; we discuss the physical device mapping in the following subsection.

{
\begin{small}
\begin{algorithm}[t]
\caption{Parallel Scheme Generation}
\label{alg:parallel_scheme}
\begin{algorithmic}[1]
\State{\textbf{Input:} LLM (in Transformer IR), Parallel Templates}
\State{\textbf{Output:} parallel\_schemes}
\State{$n$ = num\_of\_devices\_in\_cluster}
\For{$model\_DP = 1, 2,...,n$:}
{\Comment{model-level DP}}
\State{$m$ = $n \div model\_DP$}
{\Comment{$m$ = num\_repica\_devices}}
\For{$stage = 1,2,..,m$}
{\Comment{inter-layer parallelism}}
\State{$s$ = $m \div stage$}
{\Comment{$s$ = num\_stage\_devices}}
\For{cell $\in$ LLM.block}
\For{$cell\_DP = 1, 2,...,s$:}
{\Comment{cell-level DP}}
\State{$c$ = $s \div cell\_DP$}
\Comment{$c$ = num\_cell\_devices}
\State{task\_mapping = \textbf{templates(}cell, c\textbf{)}}
\State{cell\_schemes.\textbf{append(}task\_mapping\textbf{)}}
\EndFor
\EndFor
\For{$r$ in 0,..., len(LLM.block.cells):}
\State{reshard = \textbf{get\_reshard\_collective(}cell\_schemes, $r$\textbf{)}}
\State{stage\_schemes.\textbf{append(}cell\_schemes, reshard\textbf{)}}
\EndFor
\State{parallel\_schemes.\textbf{append(}stage\_schemes\textbf{)}}
\EndFor
\EndFor
\end{algorithmic}
\end{algorithm}
\end{small}
}

\subsubsection{Device Mapper}\label{sec:mapper}
The Device Mapper takes the parallel schemes and the device cluster configuration as input, and assigns logical devices cluster to the physical device cluster, producing parallel execution plans. 
While the parallel schemes are generated in a top-down approach, the Device Mapper operates in a bottom-up manner.
Specifically, the logical devices are first mapped to physical devices connected at the bottom level of the cluster, such as Level 1 in Figure \ref{fig:cluster}; if the number of logical devices exceeds the available physical devices at the current level, the mapping extends to the next higher level to include additional physical devices.
Since lower-level connections generally offer higher bandwidth (e.g., NVLink in Level 1), the Device Mapper prioritizes mapping logical devices assigned to the same cell, as fine-grained parallelism (e.g., tensor parallelism) often incurs expensive collective communication (e.g., AllReduce).
Next, the Device Mapper maps logical devices assigned to the same pipeline stage, followed by those assigned to the same model replica, progressing through increasingly coarse-grained parallelisms.
This bottom-up approach ensures that logical devices with higher communication demands are mapped to lower-level physical connections with greater bandwidth, while those with lower communication overhead are placed at higher levels.
Once the mapping is completed, we obtain multiple parallel execution plans, each describing a distinct strategy for parallelizing the input LLM on the target cluster. 
Next, these plans are evaluated using the Batching Module and the LLM Serving Simulator.



\subsection{Batching Module}\label{sec:batcher}
The Batching Module takes the parallel execution plans and a trace of requests as inputs, and simulates LLM serving system using \textit{iteration-level batching} \cite{orca}.
Request traces are often collected from online services like Azure LLM Inference traces \cite{splitwise}, and each request comprises three key attributes: context length, generation length, and arrival time.
As mentioned in Section \ref{sec:serving}, iteration-level batching greedily adds arriving requests to the processing batch whenever memory permits.
Therefore, the Batching Module tracks memory usage and request arrival times to determine whether new requests can be added to the processing batch. 
It maintains an active request list that tracks requests that are in the processing batch and the length of their generated tokens. 
When a request reaches its generation length, it is removed from the batch, freeing memory for new requests.
In each iteration, one token is generated per active request, and the active request list is sent to the LLM Serving Simulator (Section \ref{sec:simulator}) to estimate the execution time and energy consumption of that iteration.
Due to the greedy nature of iteration-level batching, the Batching Module may add requests that exceed memory capacity, as it does not pre-allocate enough memory for the KV cache of generated tokens.
In such cases, the most recently added requests and their tokens are temporarily removed to free memory for earlier requests to complete. 

\vspace{-0.1cm}
\subsection{LLM Serving Simulator}\label{sec:simulator}
The LLM Serving Simulator estimates the execution time and energy consumption for each iteration based on active requests from the Batching Module and Profiling Results. 
The Profiling Results provide execution times and energy consumptions for key operations such as multi-head attention under various configurations (e.g., number of attention heads, feature dimensions). 
If a specific data point is missing, the Simulator applies linear interpolation between the nearest profiling data points. 
We discuss the profiling process in detail in the next subsection and focus here on the Simulator's execution process.
The Simulator estimates the time to process the requests in the active request list. 
For requests in prefill stage, execution time is estimated by querying the profiling results with the full context length, as the entire sequence is processed in parallel. 
For requests in decode stage, only the current token is processed, reducing the effective context length to one. 
If there are $n$ decode requests, execution time is estimated using context length $n$, since all decode requests (each with an effective context length of one) are processed in parallel.
The Simulator estimates energy consumption in a similar fashion by querying the profiling results for the energy consumed by each operation.
To reduce simulation overhead, the Simulator leverages the repetitive structure of LLMs by simulating only a single Transformer \textit{block}, which is the smallest set of non-repetitive \textit{cells} of a LLM (Section \ref{sec:ir}). 
APEX then extrapolates the results to estimate the execution time and energy consumption of the full model.
For example, the execution time of the iteration is extrapolated by taking the maximum across all pipeline stages, since the latency is determined by the slowest pipeline stage. In contrast, energy consumption is computed as the sum across all stages, as all devices contribute to energy usage.
The Simulator also tracks key LLM serving metrics such as time to first token, time per output token, P95 latency, Model FLOPs Utilization, and Model Bandwidth Utilization; these metrics allow APEX to evaluate the parallel execution plans comprehensively and pick an optimal plan based on a user's desired metric(s).

\vspace{-0.1cm}
\subsection{Profiling Results}\label{sec:profiler}
The profiling results are collected offline and used as \textit{system data} (Figure \ref{fig:overview}) during simulation.
An Offline Profiler collects platform-specific performance data, which is used by the LLM Serving Simulator to estimate execution time and energy consumption. 
Instead of profiling each LLM exhaustively, the Profiler performs \textit{operation-level profiling}, measuring key transformer operations such as multi-head attention and general matrix multiplication. Since these operations are fundamental to transformer models, this approach allows APEX to support various LLMs without requiring per-model profiling.
The profiling is performed across various context lengths, attention heads, hidden dimensions, etc. 
Additionally, the profiler measures collective communication overheads, including AllReduce, ReduceScatter, etc., across different data transfer sizes, device counts, and number of nodes. 
By covering a wide range of computation and communication configurations, the profiling results facilitate accurate estimation of execution time and energy consumption across different LLMs, device clusters, and parallel execution plans. 
Notably, profiling is a one-time cost when porting to a new cluster and can be amortized across multiple simulations.

\section{Experimental Results}\label{sec:experiments}

\begin{table}[t]
\centering
\caption{Detail of the request traces used for evaluation}
\vspace{-0.25cm}
\label{tab:datasets}
\resizebox{\columnwidth}{!}{%
\begin{tabular}{cccc}
\toprule
Traces       & Context Lengths & Generation Lengths & \# of requests            \\ \midrule \midrule
Summarization &        2742.11$\pm$944.33         &    172.22$\pm$73.17    &    1188                            \\  \midrule
Creation      &      306.82$\pm$81.03    &   1128.34$\pm$419.64    &   512                             \\  \midrule
Chat          & 73.32$\pm$148.65  & 189.47$\pm$174.18    & 1024                  \\ \bottomrule
\end{tabular}%
}
\vspace{-0.25cm}
\end{table}

\newcolumntype{Y}{>{\centering\arraybackslash}X}
\begin{table*}[t]
\centering
\caption{APEX simulation results: End-to-end latency (seconds). APEX demonstrates its ability to identify optimal parallel execution plans, achieving superior serving performance compared to heuristic baseline plans.}
\label{tab:main_results}
\renewcommand{\arraystretch}{0.85}
\begin{tabularx}{\textwidth}{YYc|YYY}
\toprule
Traces                         & Model                          & Arrival Rate & Baseline             & Feasible Optimal       & APEX Optimal           \\ \midrule \midrule
\multirow{6}{*}{Summarization} & \multirow{2}{*}{Llama-3.1-70B} & 0.25         & 1998.82 (1$\times$) & 1340.99 (1.49$\times$)  & 1175.09 (1.70$\times$) \\ \cline{3-6} 
                               &                                & 0.5          & 1565.11 (1$\times$) & 1225.38 (1.28$\times$)  & 1140.19 (1.37$\times$) \\ \cline{2-6} 
                               & \multirow{2}{*}{Mistral-Large} & 0.25         & 2210.05 (1$\times$) & 1624.17 (1.36$\times$)   & 1514.19 (1.46$\times$)  \\ \cline{3-6} 
                               &                                & 0.5          & 1683.00 (1$\times$) & 1483.38 (1.14$\times$)   & 1420.98 (1.18$\times$)  \\ \cline{2-6} 
                               & \multirow{2}{*}{Mixtral-8x22B} & 0.25         & 2005.06 (1$\times$) & 1753.79 (1.14$\times$)  & 859.83 (2.33$\times$)  \\ \cline{3-6} 
                               &                                & 0.5          & 1618.54 (1$\times$) & 1473.79 (1.10$\times$)  & 856.84 (1.89$\times$)  \\ \hline
\multirow{6}{*}{Creation}      & \multirow{2}{*}{Llama-3.1-70B} & 0.25         & 4027.70 (1$\times$) & 2301.54 (1.75$\times$)  & 1945.75 (2.07$\times$) \\ \cline{3-6} 
                               &                                & 0.5          & 1869.45 (1$\times$) & 1639.87 (1.14$\times$)  & 1133.05 (1.65$\times$) \\ \cline{2-6}
                               & \multirow{2}{*}{Mistral-Large} & 0.25         & 3561.55 (1$\times$) & 2268.50 (1.57$\times$)  & 1496.45 (2.38$\times$) \\ \cline{3-6} 
                               &                                & 0.5          & 2560.70 (1$\times$) & 2016.30 (1.27$\times$)  & 1313.18 (1.95$\times$) \\ \cline{2-6} 
                               & \multirow{2}{*}{Mixtral-8x22B} & 0.25         & 2645.79 (1$\times$) & 2593.91  (1.02$\times$) & 1496.45 (1.77$\times$) \\ \cline{3-6} 
                               &                                & 0.5          & 1393.19 (1$\times$) & 1386.16 (1.01$\times$)  & 1035.62 (1.35$\times$) \\ \hline
\multirow{6}{*}{Chat}          & \multirow{2}{*}{Llama-3.1-70B} & 0.25         & 1622.48 (1$\times$) & 1118.85 (1.45$\times$)  & 824.43 (1.97$\times$)  \\ \cline{3-6} 
                               &                                & 0.5          & 1344.42 (1$\times$) & 1021.09 (1.32$\times$)  & 808.96 (1.66$\times$)  \\ \cline{2-6} 
                               & \multirow{2}{*}{Mistral-Large} & 0.25         & 2041.91 (1$\times$)  & 1423.44 (1.43$\times$)   & 1182.75 (1.73$\times$)  \\ \cline{3-6} 
                               &                                & 0.5          & 1594.74 (1$\times$)  & 1356.52 (1.18$\times$)   & 1102.20 (1.45$\times$)  \\ \cline{2-6} 
                               & \multirow{2}{*}{Mixtral-8x22B} & 0.25         & 1938.26 (1$\times$) & 1658.22 (1.17$\times$)  & 575.12 (3.37$\times$)   \\ \cline{3-6} 
                               &                                & 0.5          & 1507.11 (1$\times$) & 1380.34 (1.09$\times$)  & 567.53 (2.66$\times$)   \\ \bottomrule
\end{tabularx}%
\end{table*}

\subsection{Experimental Setup}\label{sec:setup}
To evaluate APEX, we choose a broad range of LLMs and datasets with distinct workloads. 
We discuss the setup details below.

\noindent \textbf{Models}: For evaluation, we choose four state-of-the-art LLMs: Llama-3.1-70B-Instruct, Llama-3.1-405B-Instruct \cite{llama3}, Mistral-Large-Instruct (123B) \cite{mistral}, and Mixtral 8x22B-Instruct \cite{mixtral}; 
This covers a myriad of LLMs in different sizes and also a Mixture-of-Expert (MoE) model.
While the models use half-precision (FP16) by default, we also test cases where the model is quantized to FP8 format.

\noindent \textbf{Datasets}:
We prepared three sets of request traces derived from distinct datasets: a paper summarization dataset \cite{arxiv_dataset}, a news abstract summarization dataset \cite{creation_dataset}, and a conversational dataset, LMSYS-Chat-1M \cite{chat_dataset}.
These datasets represent three distinct workloads.
The paper summarization \cite{arxiv_dataset} exemplifies a prefill-intensive workload, which has a long context length and short generation length.
To create a generation-intensive workload with short context and long generation lengths, we modify the news abstract summarization dataset \cite{creation_dataset} by appending the following prompt: \textit{``Generate a long story using the provided abstract."}
This prompts the LLM to produce a long story from a short abstract, resulting in a generation-intensive workload.
The LMSYS-Chat-1M \cite{chat_dataset} dataset collects real-world conversations between users and LLMs.
Most requests in this dataset feature short context and generation lengths, representing a lightweight conversational workload.
We randomly subsampled between 512 and 1188 requests from each dataset to produce our request traces.
We refer to these traces as \textit{Summarization} (paper summarization), \textit{Creation} (news generation), and \textit{Chat} (LMSYS-Chat-1M) in the following sections.
We assume a Poisson distribution~\cite{poisson} for request arrival times. 
Details of the request traces are listed in Table~\ref{tab:datasets}.

\noindent \textbf{Hardware Platform and Serving Systems}: 
We run APEX simulation using an Intel Xeon 6530 CPU and validated the simulation results against the actual performance of LLM serving on a GPU cluster.
We deployed vLLM \cite{vllm} v0.6.0 and SGLang v0.4.5 \cite{sglang} to configure LLM online servers on three GPU clusters: (1) a Single-node H100 cluster equipped with 8 Nvidia H100 SXM GPUs, each with 80 GB of GPU memory, (2) a Multi-node H100 Cluster consisting of two nodes, each with 8 Nvidia H100 SXM GPUs, and (3) a H200 Cluster (single-node) that has 8 Nvidia H200 SXM GPUs.




\subsection{Evaluation of APEX's prediction}\label{sec:plan}
In this experiment, we address the following research question: 
\textbf{Can APEX improve LLM serving performance by identifying optimal parallel execution plans?}
For evaluation, we design multiple tasks, with each task comprising an LLM, request trace, and arrival rate.
We run APEX simulation for each task and compare the end-to-end serving latency of the following three plans:

\vspace{0.1cm}
\noindent \textbf{1. Baseline plan}: We follow the commonly-used heuristics \cite{flexflow,heuristic}, which applies tensor parallelism within the same node and pipeline parallelism across nodes.

\vspace{0.1cm}
\noindent \textbf{2. APEX Optimal plan}: APEX identifies an optimal plan for each task. However, APEX's search space extends beyond the capabilities of current LLM serving systems, including advanced features like cell-level data parallelism. As a result, an execution plan may not be fully supported by existing LLM serving systems. Still, we include the results to showcase the potential performance gains that can be achieved by supporting a broader range of parallelism techniques.

\vspace{0.1cm}
\noindent \textbf{3. Feasible Optimal plan}: This is the optimal plan identified by APEX under the constraint that only parallelism techniques supported by current LLM serving systems are used. Due to the reduced search space, the feasible optimal plan may achieve lower performance compared to the unconstrained APEX Optimal plan.

\subsubsection{Single-Node H100 cluster}\label{sec:single_node}
We first evaluate APEX on a single-node cluster with 8 H100 GPUs and show the experimental results in Table \ref{tab:main_results}. 
We experiment with the three request traces and three LLMs.
We also conduct experiments using two different arrival rates for the requests, assuming a Poisson distribution~\cite{poisson}.
To explore different quantization formats, we quantize the Mistral-Large model using FP8 for the KV cache, as well as the weights and activations (i.e., W8A8).
APEX identifies optimal parallel execution plans for both dense and sparse LLMs (i.e, MoE models), and the Feasible Optimal plans consistently deliver performance improvement over the baseline, achieving up to 1.75$\times$ speedup in terms of end-to-end serving latency.
Furthermore, assuming cell-level data parallelism is supported, APEX Optimal Plans achieve speedups of up to 3.37$\times$.
In this experiment, we demonstrate APEX is able to identify an optimal parallel execution plan that improves LLM serving latency under various LLMs and request traces.
The identified optimal plans consist of various combinations of data, pipeline, and tensor parallelism, effectively balancing the trade-offs of various factors (e.g., compute, memory, network traffic) to outperform the baseline plan, which only relies on tensor parallelism.
While the optimal plans are different from task to task, we observe that incorporating data parallelism (DP) often yields performance benefits. 
Many of the identified optimal plans set the degree of DP to 2. 
Existing LLM serving systems often overlook DP, assuming it is prohibitively memory-intensive. 
Instead, our results demonstrate that trading memory efficiency for reduced communication overhead can lead to performance improvements.
We also observe that the Feasible Optimal plan of Mixtral-8x22B shows only marginal improvement over the baseline on the Creation trace.
This is due to the dominating role of memory utilization in this setup,  leaving fewer trade-offs available for performance improvement. Specifically, Mixtral-8x22B, with its 141B parameters, heavily utilizes memory resources, and the Creation traces’s generation-intensive nature further increases memory demands due to extensive KV caching. 
Under these conditions, the baseline plan, which adopts pure tensor parallelism, performs well owing to its high memory efficiency.
Nevertheless, the APEX Optimal plan still manages to deliver a 1.77$\times$ speedup by enabling cell-level DP.

\begin{table}[t]
\centering
\caption{End-to-End latency (sec.). APEX identifies optimal execution plans for different clusters}
\vspace{-0.1cm}
\label{tab:multi_nodes}
\renewcommand{\arraystretch}{0.8}
\begin{tabular}{cc|cc}
\toprule
                          Model \& Cluster                                                  & Traces   & Baseline & APEX Optimal    \\ \midrule \midrule
\multirow{3}{*}{\begin{tabular}[c]{@{}c@{}}Llama-3.1-405B\\ (Multi-Node H100)\end{tabular}} & Summ.    & 4077.88  & 2519.12 (1.62$\times$) \\ \cline{2-4} 
                                                                                            & Creation & 2314.25  & 1229.14 (1.88$\times$) \\ \cline{2-4} 
                                                                                            & Chat     & 2645.89  & 1710.10 (1.55$\times$) \\ \midrule
\multirow{3}{*}{\begin{tabular}[c]{@{}c@{}}Llama-3.1-70B\\ (H200)\end{tabular}}             & Summ.    & 1433.21  & 897.95 (1.60$\times$)  \\ \cline{2-4} 
                                                                                            & Creation & 1128.66  & 667.85 (1.69$\times$)  \\ \cline{2-4} 
                                                                                            & Chat     & 1189.00     & 536.74 (2.22$\times$)  \\ \midrule
\multirow{3}{*}{\begin{tabular}[c]{@{}c@{}}Mistral-Large\\ (H200)\end{tabular}}             & Summ.    & 1595.24  & 1174.79 (1.36$\times$) \\ \cline{2-4} 
                                                                                            & Creation & 1389.19  & 776.08 (1.79$\times$)  \\ \cline{2-4} 
                                                                                            & Chat     & 1515.41  & 961.59 (1.56$\times$)  \\ \midrule
\multirow{3}{*}{\begin{tabular}[c]{@{}c@{}}Mixtral-8x22B\\ (H200)\end{tabular}}             & Summ.    & 1540.42  & 706.06 (2.18$\times$)  \\ \cline{2-4} 
                                                                                            & Creation & 1380.75  & 993.34 (1.39$\times$) \\ \cline{2-4} 
                                                                                            & Chat     & 1380.76  & 500.32 (2.76$\times$)  \\ \bottomrule

\end{tabular}
\vspace{-0.3cm}
\end{table}

\subsubsection{Multi-Node H100 Cluster}
Next, we evaluate APEX on a two-node cluster, each node contains 8 NVIDIA H100 SXM GPUs, identical to the setup in Section \ref{sec:single_node}. 
We use Llama-3.1-405B, one of the largest publicly available LLMs for evaluation on this cluster.  
Results are presented in the first row of Table \ref{tab:multi_nodes}.
Due to space constraints, we report only the Baseline Plan and APEX Optimal Plan under an arrival rate of 0.5. Experimental results show that APEX consistently identifies execution plans that outperform the baseline, achieving speedups ranging from 1.55$\times$ to 1.88$\times$.
These findings demonstrate APEX's ability to scale effectively to multi-node clusters and extremely large models, continuing to identify performant execution plans under increased system complexity.

\subsubsection{H200 Cluster}
We also evaluate APEX on a cluster with 8 NVIDIA H200 GPUs. Compared to the H100 (80 GB HBM), the H200 offers the same compute and peak performance but with increased memory capacity (141 GB HBM). This expanded memory enables the H200 cluster to explore a larger design space, allowing for more flexible trade-offs between parallelism strategies.
As shown in last three rows of Table \ref{tab:multi_nodes}, APEX finds simulation plans that lead to up to 2.76$\times$ speedup compared to the baseline plan. 
Notably, APEX recommends different optimal plans for the H200 and H100 clusters under the same experimental setup. For instance, when serving the Summarization trace with the Mistral-Large model, APEX suggests 2-way data parallelism for H100, and 4-way data parallelism plan for H200. 
While the H200 plan incurs higher memory overhead, it reduces communication overhead, making the trade-off favorable due to the H200's larger memory capacity.

\subsubsection{Optimize Towards Energy Efficiency}
For service providers, latency is not the only metric that matters. Rather than minimizing inference time at all costs, providers are typically required to meet specific service-level objectives (SLOs), allowing room to optimize other metrics as long as the SLOs are satisfied.
One critical metric is energy consumption, which directly impacts both operational costs and carbon emissions. 
APEX supports optimizing for energy consumption by profiling not only execution time but also the energy usage of operations.
We demonstrate this capability in Table~\ref{tab:energy}, which presents results for serving a Llama-3.1-70B model on a cluster with 8 H100 GPUs. The first rows report latency-optimal plans (i.e., the APEX Optimal Plans in Table~\ref{tab:main_results}), which generally yield better energy efficiency than the baseline plans due to shorter runtime. 
However, latency-optimal plans are not necessarily energy-optimal. 
We observe that energy efficiency varies across operations of different input sizes, and some operations consume disproportionately more energy; investigating the underlying cause of this phenomenon is beyond the scope of this work, and we instead focus on leveraging the collected profiling results to identify energy-efficient execution plans.
APEX evaluates the energy consumption across various parallel execution plans and identifies the energy-optimal plan.
Reported in the second rows of Table~\ref{tab:energy}, energy-optimal plans reduce energy consumption by up to 19\% compared to latency-optimal plans.
For systems with more relaxed SLOs, service providers may further reduce energy consumption by lowering GPU frequency. 
In the third rows of Table~\ref{tab:energy}, APEX identifies optimal plans under 0.8 GHz of GPU frequency, reducing energy consumption by 45\% and 41\% on the Summarization and Creation traces, respectively. 
While this results in increased TTFT (up to 1.9$\times$) and TPOT (up to 1.8$\times$), such increases may still be acceptable in practice, for instance, when the TPOT requirement is 50 ms. These results show that APEX enables service providers to make informed trade-offs between latency and energy efficiency.

\begin{table}[]
\caption{APEX identifies plans optimized toward energy}
\vspace{-0.15cm}
\label{tab:energy}
\begin{tabularx}{\columnwidth}{YcYYY}
\toprule
Traces                    & \begin{tabular}[c]{@{}c@{}}Exe. Plans \\ (GPU Freq., GHz)\end{tabular} & \begin{tabular}[c]{@{}c@{}}Energy \\ (KJ)\end{tabular} & \begin{tabular}[c]{@{}c@{}}TTFT\\ (ms)\end{tabular} & \begin{tabular}[c]{@{}c@{}}TPOT \\ (ms)\end{tabular} \\ \midrule \midrule
\multirow{3}{*}{Summ.}    & Latency opt. (2.0)                                             & 4.095                                                       & 788.67                                              & 32.06                                                \\ \cline{2-5} 
                          & Energy opt. (2.0)                                              & 3.293                                                       & 992.69                                              & 26.89                                                \\ \cline{2-5} 
                          & Energy opt. (0.8)                                              & 2.265                                                       & 1485.94                                             & 33.62                                                \\ \midrule
\multirow{3}{*}{Creation} & Latency opt. (2.0)                                             & 8.388                                                       & 44.67                                               & 18.83                                                \\ \cline{2-5} 
                          & Energy opt. (2.0)                                              & 8.277                                                       & 54.59                                               & 20.63                                                \\ \cline{2-5} 
                          & Energy opt. (0.8)                                              & 5.016                                                       & 75.70                                                & 34.59                                                \\ \bottomrule
\end{tabularx}
\vspace{-0.35cm}
\end{table}





\subsection{Evaluation of Simulation Fidelity}\label{sec:fidelity}
\begin{figure}[t]
    \centering
    \includegraphics[width=8.5cm]{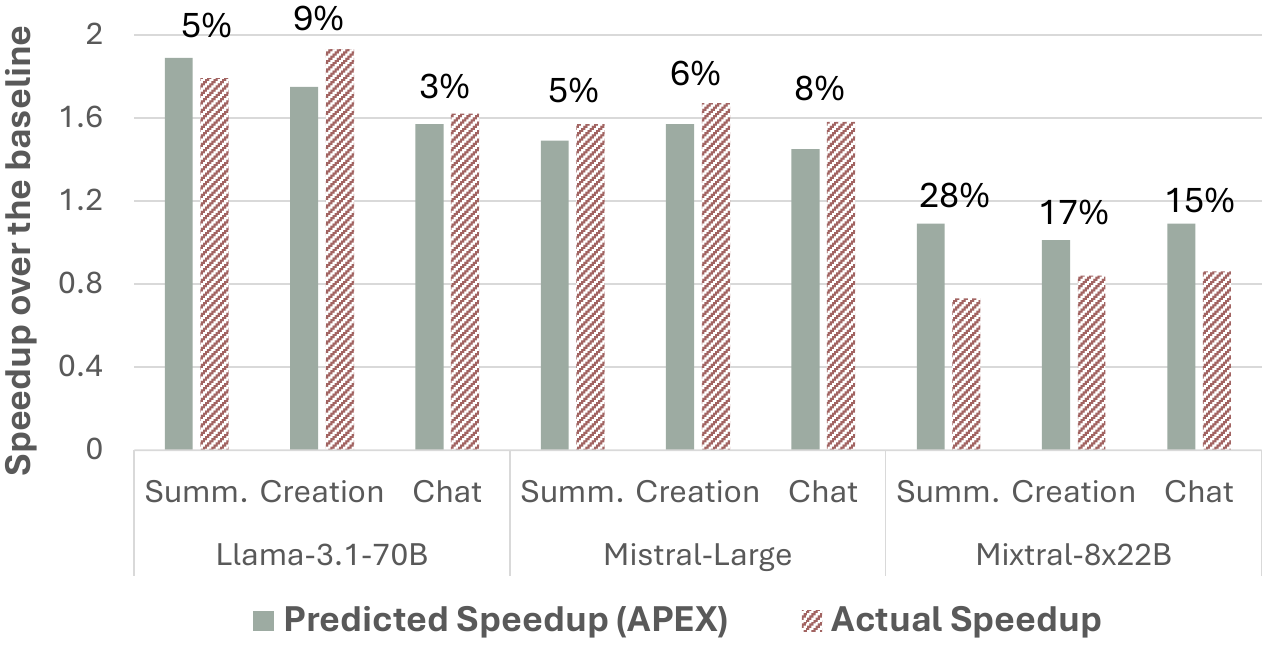}
    \vspace{-0.5cm}
    \caption{Speedup predicted by APEX vs. actual speedup achieved. APEX accurately predicts the speedup of adopting an optimal execution plan over the baseline plan.}
    \vspace{-0.5cm}
     \label{fig:fidelity1}
\end{figure}

\begin{figure}[t]
    \centering
    \includegraphics[width=8.5cm]{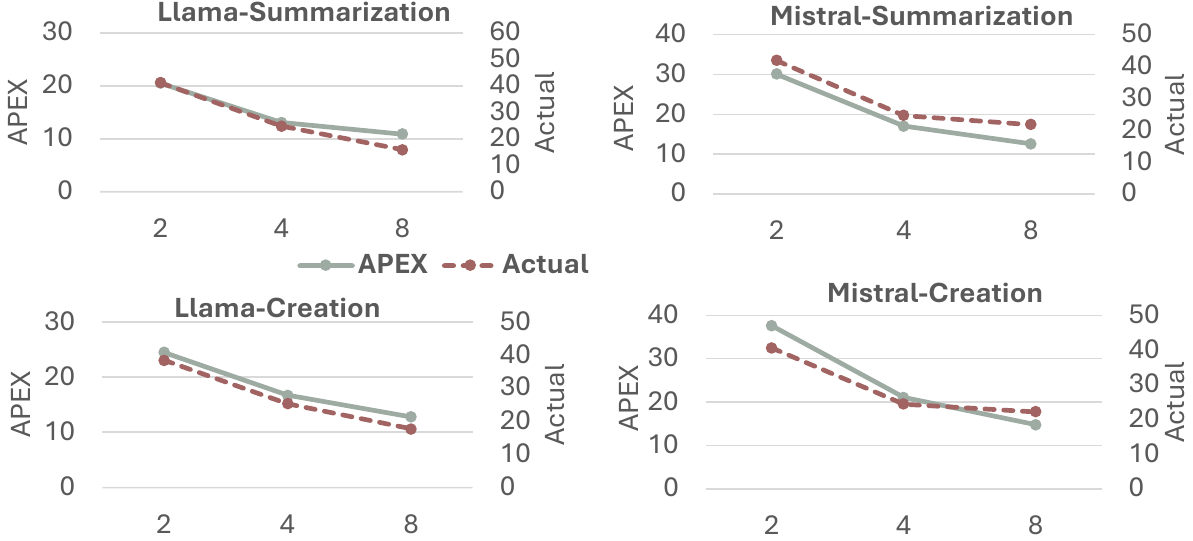}
    \caption{APEX accurately predicts the scalability trend across varying numbers of GPUs, as measured by TPOT (ms)}
    \vspace{-0.6cm}
     \label{fig:fidelity2}
\end{figure}

In this experiment, we address the following research question: 
\textbf{Can APEX provide high-fidelity simulation for LLM serving?}
While we demonstrate in Section~\ref{sec:plan} that APEX can identify optimal parallel execution plans that enhance LLM serving performance, it is also essential to verify the fidelity of its simulation results.
We evaluate the simulation fidelity from two perspectives.
First, under a fixed number of devices, we assess whether APEX can accurately predict the performance improvement of an optimal plan compared to a baseline plan. 
Second, for a given execution plan, we evaluate whether APEX can reliably predict performance improvements when scaling across different numbers of devices.
We show the experimental results in Figure \ref{fig:fidelity1} and Figure \ref{fig:fidelity2}.

For the first experiment set, we compare the predicted speedup of Feasible Optimal plans with the actual speedup achieved on the device cluster (Figure \ref{fig:fidelity1}).
We use vLLM~\cite{vllm} to serve Llama-3.1-70B and Mistral-Large, and use SGLang~\cite{sglang} for Mixtral-8x22B as vLLM does not support expert parallelism (EP) at the time of our experiments.
APEX achieves accurate predictions, with an average relative error of 10.7\%.
The main  discrepancy lies in the predictions for Mixtral-8x22B. 
APEX estimates that applying EP would outperform tensor parallelism (TP) due to the lower overhead of All-to-All communication in EP, compared to the AllReduce operations required in TP. 
However, when serving with SGLang, we observe that TP consistently outperforms EP across experiments. 
A potential reason is that the current implementation of EP may not be as well-optimized as TP, leading to higher-than-expected overheads during execution.

For the second experiment set, we evaluate the time per output token (TPOT) across different numbers of GPUs using tensor parallelism.
We show the results in Figure \ref{fig:fidelity2}.
The APEX prediction results are shown as green solid lines, while the actual results are represented by red dotted lines. 
The high similarity between the two lines across all cases demonstrates that APEX accurately captures the scalability trend when scaling from 2 GPUs to 8 GPUs across various models and request traces.
The estimated TPOT (ms) of APEX is plotted on the left-hand Y-axis, and the actual TPOT (ms) is plotted on the right-hand Y-axis. 
The actual measured TPOT is consistently higher than the predicted value, as evident from the larger values on the right-hand Y-axis compared to the left-hand Y-axis.
This is because APEX focuses on the overhead of key operations, such as attention and feedforward, while omitting the overhead of other operations.
Nevertheless, APEX accurately captures the relative performance differences between various parallel execution plans and degrees of parallelism, as demonstrated in Figure \ref{fig:fidelity1} and \ref{fig:fidelity2}.
This capability enables APEX to effectively evaluate and compare different execution plans, thereby identifying an optimal parallel execution plan for LLM serving.

\subsection{Evaluation of Efficiency}
We evaluate the efficiency of APEX by comparing the APEX simulation time with the actual execution time on the device cluster.
Evaluating all parallel execution plans across the setups in Table \ref{tab:main_results} (i.e., different models, traces, arrival rates) takes approximately 160 hours on 8 H100 GPUs. 
In contrast, the same evaluation is completed in less than 2.5 hours using APEX on a CPU, making it 71$\times$ faster than the actual deployment.
From a cost perspective, running the actual implementation would cost approximately $\$$8,889 (based on the Azure NC40ads H100 cluster pricing), whereas the APEX simulation costs only $\$$7.20 (assuming an Azure D64s v6 CPU node). This translates to a 1234.5$\times$ cost reduction, demonstrating the significant time efficiency and cost-effectiveness of APEX.
Obtaining the operation-level profiling results to set up APEX takes approximately 40 GPU hours.
Yet, this is a one-time cost that can be amortized for the same type of device cluster.

\begin{figure}[t]
    \centering
    \includegraphics[width=7.5cm]{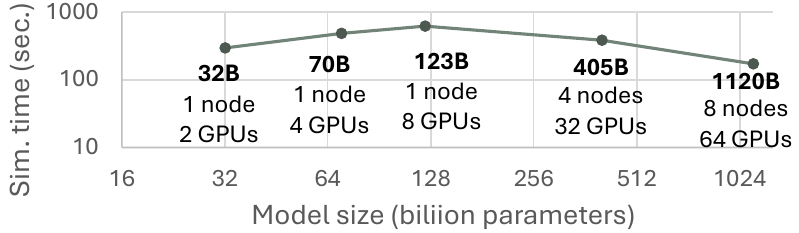}
    \vspace{-0.2cm}
    \caption{APEX maintains similar simulation overhead when scaling from billion-scale to trillion-scale models}
    \vspace{-0.6cm}
     \label{fig:scalability}
\end{figure} 

We also evaluate the simulation overhead when scaling to larger LLMs on larger device clusters.
We use the following example models: Qwen2.5-32B, Llama-3.1-70B, Mistral-Large, and Llama-3.1-405B.
We synthesize a trillion-scale model by scaling the Llama-3.1-70B model 16 times; this is done by modifying values in the LLM configuration file~\cite{config}.
As shown in Figure \ref{fig:scalability}, APEX demonstrates high scalability, maintaining a similar simulation overhead when scaling from a 32B to a trillion-scale model.
This efficiency is achieved by leveraging the canonical representation of Transformer IR (Section \ref{sec:ir}) and the repetitive structure of the transformer architecture to reduce the design space.
This scalability highlights APEX's potential for future applications, as model sizes continue to grow and the cost of actual deployment become more expensive.

\subsection{Evaluation of Extensibility}\label{sec:extensibility}
While APEX supports a wide range of models, devices, etc., the simulator may still become outdated as new advancements emerge. 
To address this, APEX is designed for extensibility, enabling easy adaptation to support the latest developments.
We evaluate APEX's extensibility by measuring the overheads to implement a new feature.
We use two metrics: (1) programming overhead, measured in lines of code required for a feature extension, and (2) implementation time overhead, measured in hours, including time to write and execute the code or scripts.
We reported the overhead of various types of extensions in Table \ref{tab:extension} and discuss the details below.

\noindent \textbf{Extending to New Models} 
APEX can effortlessly support a new LLM without additional programming or implementation time by requiring only a configuration file~\cite{config} of the model. 
However, for LLMs containing unknown transformer cells (Section \ref{sec:ir}), such as a novel feedforward network, additional effort is needed. 
The primary work involves implementing the Parallel Template (Section \ref{sec:template}) for the new cells. 
For evaluation, we measured the effort to support SwiGLU cells \cite{swiglu} used in the Llama-3.1 models.

\noindent \textbf{Extending to New Device Cluster}
APEX can easily adapt to new device clusters by providing the device name, memory capacity, and interconnection network (e.g., NVLink, PCIe).
We evaluated the effort required to support a new GPU model.
The implementation time overhead mainly involves running profiling scripts, which take 6 to 8 hours. 
While extending to new devices requires a relatively long profiling time, this is a one-time effort that can be amortized.

\noindent \textbf{Extending to New Batching Mechanism}
APEX also supports new batching mechanisms.
While this is not as straightforward as supporting a new model or a new device cluster, it can be done by modifying the Batching Module (Section~\ref{sec:batcher}).
APEX adopts contiguous batching \cite{vllm,sglang} by default.
We evaluate the overhead of extending to support a Sarathi-Serve-style \cite{Sarathi-Serve} batching, which performs \textit{chunked prefill} to better interleave prefill and decode requests.
This extension is done by adding a new variable, \textit{chunk size}, to the Batching Module and a counter to each request to ensure all prefill chunks are completed before moving to the decode stage.

\begin{table}[t]
\centering
\caption{Evaluating the overhead of extending APEX}
\vspace{-0.2cm}
\label{tab:extension}
\renewcommand{\arraystretch}{0.9}
\resizebox{\columnwidth}{!}{%
\begin{tabular}{c|cc}
\toprule
\multicolumn{1}{c}{\multirow{2}{*}{Extension Type}} & \multicolumn{1}{c}{\multirow{2}{*}{\begin{tabular}[c]{@{}c@{}}Programming  Overhead \\ (Lines of Code)\end{tabular}}} & \multicolumn{1}{c}{\multirow{2}{*}{\begin{tabular}[c]{@{}c@{}}Implementation Time \\ Overhead (Hour)\end{tabular}}} \\
\multicolumn{1}{c}{}                                & \multicolumn{1}{c}{}                                                                                                  & \multicolumn{1}{c}{}                                                                                                \\ \midrule \midrule
LLM                            & 0                                                                                                                     & $\sim$0                                                                                                             \\ \hline
LLM (w/ unknown cells)                          & 50 - 150                                                                                                              & 1-2                                                                                                                 \\ \hline
Device cluster                                      & $\sim$20                                                                                                              & 6-8                                                                                                                 \\ \hline
Batching Mechansim                                  & $\sim$100                                                                                                             & 1-2                                                                                                                 \\ \hline
Parallelism                                         & 50 - 200                                                                                                              & 1-2                                                                                                                    \\   \bottomrule                                                                                                       
\end{tabular}%
}
\vspace{-0.5cm}
\end{table}

\noindent \textbf{Extending to New Parallelism}
The overhead of extending to new parallelisms depends on the specific parallelism being added. For example, supporting a new intra-layer parallelism is analogous to supporting a new transformer cell, as it requires adding a new Parallel Template. Similarly, Fully Sharded Data Parallelism (FSDP) can be integrated into APEX by extending the existing DP implementation, incorporating additional collective communications and adjusting memory usage for the sharded model storage.




\subsection{Beyond Identifying Optimal Execution Plan}
\begin{figure}[t]
    \centering
    \includegraphics[width=8.2cm]{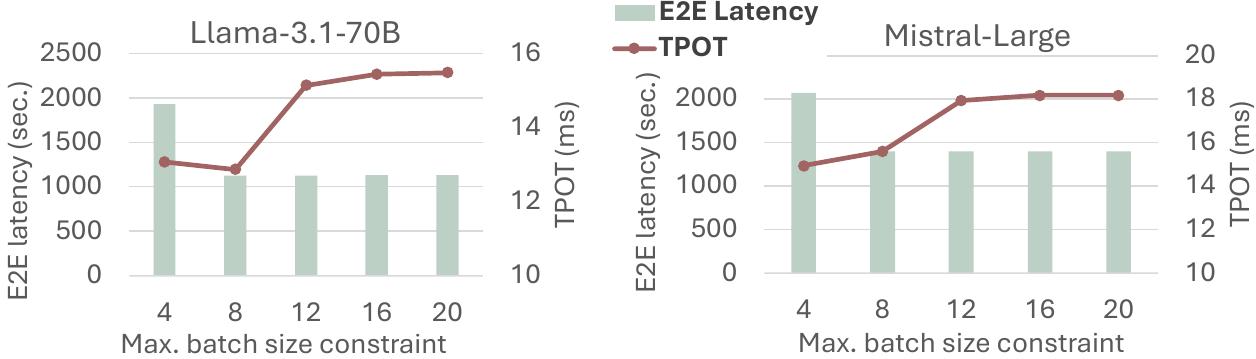}
    \vspace{-0.1cm}
    \caption{APEX can simulate with various maximum batch size constraints and suggest adjustments to meet SLOs.}
     \label{fig:slo}
     \vspace{-0.4cm}
\end{figure}

The high simulation fidelity makes APEX applicable to other use cases beyond identifing optimal execution plans. 
For example, APEX can help service providers meet various service-level objectives (SLOs), such as maintaining a target time per output token (TPOT). 
A common strategy to meet latency requirements is to adjust the batch size by setting a maximum batch size constraint.
APEX begins by simulating a subset of requests to determine an upper bound for the batch size, denoted as $m$.
It then divides this upper bound into $n$ segments and simulates the request traces with various batch size constraints, ranging from $\frac{1\times m}{n}$ to $\frac{n\times m}{n}$, where $n$ is determined heuristically.
Figure \ref{fig:slo} illustrates two examples using the Llama-3.1-70B and Mistral-Large models on the Creation trace.
Suppose a service provider aims to reduce TPOT to meet an SLO. Based on APEX's results, they can estimate the required adjustment to the batch size constraint. For instance, reducing the maximum batch size from 16 to 8 yields a TPOT improvement of 18\% for Llama and 14\% for Mistral.
However, overly restricting the maximum batch size may negatively affect performance; APEX also captures this trade-off, as shown in Figure~\ref{fig:slo}, where setting the batch size constraint to 4 leads to degraded end-to-end serving latency.



\section{Related Work}\label{sec:related_work}

\noindent \textbf{LLM Serving Systems:}
With the emergence of LLMs, numerous LLM serving systems have been proposed \cite{orca, vllm,Sarathi-Serve,distserve,tensorRT,splitwise}.
Each system introduces innovations to address key challenges in LLM serving.
Orca \cite{orca} introduced iteration-level batching, significantly improving serving throughput. This technique has since become a standard in LLM serving systems.
vLLM \cite{vllm} proposed PagedAttention, an efficient method to manage KV cache memory usage, enabling more concurrent request batching.
DistServe and Splitwise \cite{distserve,splitwise} disaggregate the prefill and decode stages, processing stage separately (e.g., using different clusters, applying different optimizations) for better serving performance.
While these systems allow users to manually specify the degree of parallelisms, they do not provide guidance on determining these configurations.
APEX complements these works by automatically identifying optimal plans, which can be used to configure these serving systems.

\noindent \textbf{Configuration Searching Tools:}
LLM-Pilot \cite{LLM_pilot} proposes a predictive model to help users select the most cost-effective hardware for serving LLMs.
DynamoLLM \cite{dyanmoLLM} proposes a framework to automatically reconfigure the cluster to optimize the energy efficiency of LLM serving.
However, both LLM-Pilot and DynamoLLM only consider tensor parallelism and can be further enhanced by incorporating hybrid parallelisms, as supported by APEX.
Calculon \cite{calculon} introduces a performance model to identify optimal execution plans for LLM training, while vTrain \cite{vTrain} and ASTRA-sim \cite{ASTRA-SIM} adopt simulator-based approaches to achieve the same goal.
Nevertheless, these solutions are tailored to training workloads and do not address the unique challenges of LLM serving, particularly the dynamism introduced by iteration-level batching.
LLMServingSim \cite{llmservingsim} and Vidur \cite{vidur} are the few frameworks designed for LLM serving. 
Yet, LLMServingSim primarily focuses on NPUs and Processing-in-Memory architectures, and it relies on their corresponding hardware simulators to estimate execution times. 
In contrast, APEX utilizes operation-level profiling data to estimate execution time and does not rely on other hardware simulators. 
Vidur requires a model onboarding step before simulation, which involves parsing the operations within the target LLM and profiling them. 
APEX bypasses this requirement by capturing the key operations of LLMs and utilizing pre-collected profiling results, allowing simulations to start immediately with amortized profiling costs. 
In addition, neither LLMServingSim nor Vidur supports energy-aware optimization, a critical requirement for LLM service providers. They also lack support for emerging LLM serving needs, such as Mixture-of-Experts models, quantization formats, and multi-node clusters.

\section{Conlusion}
In this work, we developed APEX, an extensible and dynamism-aware simulator for LLM serving. We evaluated APEX across various LLMs and distinct workloads, demonstrating its high-fidelity simulation with only a 10.7\% relative error on the average.
APEX successfully identified latency-optimal execution plans that achieve up to 3.37$\times$ speedup compared to heuristic plans, and also energy-optimal plans that reduce energy consumption by up to 45\% compared to latency-optimal plans.
The simulation proved highly efficient, providing a 71$\times$ speedup and 1234$\times$ greater cost-effectiveness than deployment on actual hardware. 
APEX maintained a similar simulation overhead when scaling from billion-scale to trillion-scale models.
While APEX already supports a broad range of LLMs, we further demonstrated its ease of extensibility to accommodate new models, device clusters, batching algorithms, and parallelisms. 
In the future, we plan to extend APEX to support multimodal LLMs, which will involve incorporating parallel execution plans for encoders handling modalities such as vision and audio.



\bibliographystyle{ACM-Reference-Format}
\bibliography{ref}

\appendix

\end{document}